# Asporin Expression Is Highly Regulated in Human Chondrocytes

*Elise Duval,[1] Nicolas Bigot,[1] Magalie Hervieu,[1] Ikuyo Kou,[2] Sylvain Leclercq,[1,3] Philippe Galéra,[1] Karim Boumediene,[1*] and Catherine Baugé[1*]*

[1]Univ Caen, EA3214 Matrice Extracellulaire et Pathologie, Caen, France; [2]Laboratory for Bone and Joint Diseases, Center for Genomic Medicine, University of Tokyo, Minato-ku, Tokyo, Japan; and the [3]Department of Orthopedic Surgery, Saint-Martin Private Clinic, Caen, France

A significant association between a polymorphism in the D repeat of the gene encoding asporin and osteoarthritis, the most frequent of articular diseases, has been recently reported. The goal of the present study was to investigate the expression of this new class I small leucine-rich proteoglycan (SLRP) in human articular chondrocytes. First, we studied the modulation of asporin (ASPN) expression by cytokines by Western blot and reverse transcription–polymerase chain reaction. Interleukin-1β and tumor necrosis factor-α downregulated ASPN, whereas transforming growth factor-β1 (when incubated in a serum-free medium) upregulated it. Similarly to proinflammatory cytokines, chondrocyte dedifferentiation induced by a successive passages of cells was accompanied by a decreased asporin expression, whereas their redifferentiation by three-dimensional culture restored its expression. Finally, we found an important role of the transcription factor Sp1 in the regulation of ASPN expression. Sp1 ectopic expression increased ASPN mRNA level and promoter activity. In addition, using gene reporter assay and electrophoretic mobility shift assay, we showed that Sp1 mediated its effect through a region located between –473 and –140 bp upstream of the transcription start site in *ASPN* gene. In conclusion, this report is the first study on the regulation of asporin expression by different cytokines in human articular chondrocytes. Our data indicate that the expression of this gene is finely regulated in cartilage and suggest a major role of Sp1.

## INTRODUCTION

Osteoarthritis (OA), the most prevalent form of skeletal disease, represents a leading cause of disability after middle age. This disease is characterized by the degeneration of joint cartilage in the knee, hip and hand. Whereas it is extremely common, the details of its etiology and pathogenesis remain unclear. Recently, a genetic association was reported between the cartilage extracellular matrix protein asporin (ASPN) and OA. It was demonstrated that OA susceptibility is affected by the number of aspartic acid (D) residues in the amino-terminal extremity of the asporin protein.

Asporin, also called periodontal ligament–associated protein-1 (PLAP-1), is a new member of the family of small leucine-rich proteoglycans (SLRPs) (1–3). The SLRP family, which composes a major noncollagen component of the extracellular matrix, consists of 13 known members that can be divided into three distinct subfamilies on the basis of their genomic organization, amino acid sequence similarity and structure (1). Asporin belongs to the group of class I SLRPs that also includes decorin (*DCN*) and biglycan (*BGN*) (2,3). Unlike other family members, asporin lacks a glycosaminoglycan attachment site and hence is not a "strict" proteoglycan.

In osteoarthritic cartilage, there is likely a functional imbalance between cytokines that promote degradation of the matrix, such as interleukin (IL)-1β or tumor necrosis factor (TNF)-α, and those that favor its repair, such as transforming growth factor (TGF)-β. Thus, IL-1β and TNFα induce the expression of metalloproteases in chondrocytes and subsequently the destruction of the cartilage. Additionally, these cytokines are able to inhibit proteoglycan and collagen synthesis. On the other hand, TGFβ may take part in the repair potentiality of cartilage by stimulating the biosynthesis of matrix components (collagens and proteoglycans). It also inhibits the production of metalloproteases and enhances the expression of their tissue inhibitors, the tissue inhibitor of metalloproteinase (TIMP). However, to our knowledge, their effect on ASPN expression has not been described until now.

---

Recent studies have enlightened interaction between ASPN and TGFβ1/bone morphogenetic protein-2 (BMP-2). ASPN directly binds to TGFβ1 and inhibits Smad signaling in chondrocytes (4), leading to the suppression of TGFβ-mediated expression of the aggrecan and collagen type II α1 genes and reduced proteoglycan accumulation. The effect on TGFβ activity was allele specific, with the D14 allele resulting in greater inhibition than other alleles. Similarly, asporin decreases BMP-2 activity by interaction with this factor and subsequently is a negative regulator of osteoblast differentiation and mineralization (5). In addition, asporin was reported to antagonize the TGFβ by physical interaction with this factor. Furthermore, the three isoforms of TGFβ are able to induce ASPN expression. However, while it is interesting, no other reports have been published about regulation of ASPN expression.

Therefore, in this study, we examined *ASPN* expression in human differentiated or dedifferentiated articular chondrocytes and its regulation by the major factors involved in OA, namely the proinflammatory cytokines IL-1β and TNFα, and TGFβ. Among the three TGFβ isoforms present in mammals, we chose to study only the effect of TGFβ1. Indeed, it is the most abundant isoform in articular cartilage, and its expression is the most affected during OA process (6). We also examined the role of Sp1, a transcription factor able to regulate others members of class I SLRPs.

## MATERIALS AND METHODS

### Cell Culture

Human articular chondrocytes (HACs) were prepared from femoral head. All donors (aged between 50 and 83 years, with a medium age of 71 years) signed agreement forms before the surgery, according to local legislations. Cells were isolated and cultured as previously described (7). Cartilage samples were cut into small slices, and chondrocytes were isolated by sequential digestion by type XIV protease (Sigma-Aldrich, St. Quentin Fallavier, France) and then type I collagenase (from *Clostridium histolyticum*; Invitrogen, Cergy-Pontoise, France). The cell suspension was filtered, seeded in plastic vessels at a density of $4 \times 10^4$ cells/cm$^2$ and cultured in Dulbecco's modified Eagle's medium (DMEM) supplemented with 10% fetal calf serum (FCS; Invitrogen), 100 IU/mL penicillin, 100 μg/mL streptomycin and 0.25 μg/mL Fungizone, in an atmosphere of 5% $CO_2$. During expansion, the medium was changed twice a week. HACs were used as primary cultures (no passage) and treated with human recombinant IL-1β (Sigma-Aldrich, St. Quentin Fallavier, France), TNFα or TGFβ1 (R&D Systems, Lille, France).

For dedifferentiation, confluent cells were harvested by trypsinization (0.25% trypsin/1 mmol/L EDTA; Invitrogen), counted and seeded again at $4 \times 10^4$ cells/cm$^2$. HACs were used in two passages, since previous experiments in the laboratory showed that two passages are sufficient to induce human chondrocyte dedifferentiation (8). For redifferentiation studies, chondrocytes, which had undergone three passages, were encapsulated in alginate beads as previously described (9). Briefly, the cells were suspended in alginate at the density of $2.5 \times 10^6$ cells/mL and then slowly dropped into a 100 mmol/L $CaCl_2$ solution. After instantaneous gelation, the beads were allowed to polymerize further for a period of 10 min in $CaCl_2$ solution. After three washes in phosphate-buffered saline (PBS), they were finally placed in DMEM supplemented with 10% FCS. After 3 wks, the beads were rinsed with PBS and dissolved in a solution com-posed of 55 mmol/L sodium citrate and 25 mmol/L EDTA. The suspension was incubated at 37°C until the beads were completely dissolved (approximately 10 min). Centrifugation of the suspension at 800*g* for 10 min allows separation of the cells from their alginate matrix. The dedifferentiation and redifferentiation of HACs after passages and culture in alginate beads were controlled by analysis of the expression of type I and II collagen.

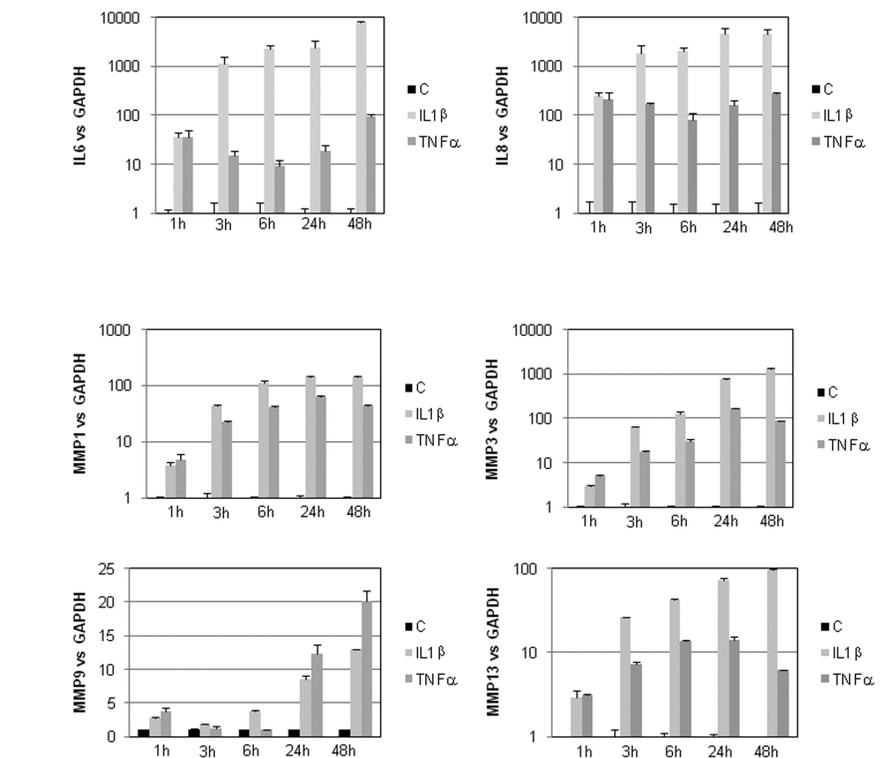

**Figure 1.** IL-1β and TNFα induces MMPs and IL-6 and IL-8 expression. Primary cultures of HACs (P0 (no passage)) were cultured for 6–7 d in DMEM + 10% FCS and then maintained for 24 h in DMEM + 2% FCS before adding 1 ng/mL IL-1β or TNFα. They were incubated with these cytokines for the indicated times. IL-6, IL-8, MMP1, MMP3, MMP8 and MMP13 mRNA were defined by RT-PCR. Histograms represent modulation of mRNA level after normalization to GAPDH signal. Values are the mean and SD of triplicate experiments. C, control.

### RNA Extraction and Real-Time Reverse Transcription–Polymerase Chain Reaction

Total RNA from primary HAC cultures were extracted using Trizol (Invitrogen by Fisher Bioblock Scientific, Illkirch, France). After extraction, 1 μg of DNase-I–treated RNA was reverse transcribed into cDNA in the presence of oligodT and Moloney murine leukemia virus reverse transcriptase (Invitrogen). The reaction was carried out at 37°C for 1 h followed by a further 10-min step at 95°C. Amplification of the generated cDNA was performed by real-time PCR in an Applied Biosystems SDS7000 apparatus with appropriate primers designed with Primer Express software. The relative mRNA level was calculated with the $2^{-\Delta\Delta CT}$ method.

### Protein Extraction and Western Blot

Cells were rinsed and scrapped in radioimmunoprecipitation assay (RIPA) lysis buffer supplemented with protease inhibitors. The extracts (30-μg proteins) were subjected to fractionation in sodium dodecyl sulfate–polyacrylamide gel electrophoresis (SDS-PAGE), transferred to polyfluorure de vinylidène (PVDF) membranes (Amersham Biosciences, Orsay, France) and reacted with goat antiasporin polyclonal antibody (ab31303; Abcam, Cambridge, UK). Subsequently, membranes were incubated with antigoat secondary peroxidase-conjugated antibody (Santa Cruz Biotechnology from tebu-bio). The signals were revealed with SuperSignal West Pico Chemiluminescent Substrate (Pierce Perbio Science, Brébières, France) and exposed to X-ray film. The membranes were also reacted with anti–β-actin (Santa Cruz Biotechnology from tebu-bio) to verify equal loading. Densitometric analyses were performed with ImageJ software, and histograms represent the ratio between ASPN and actin.

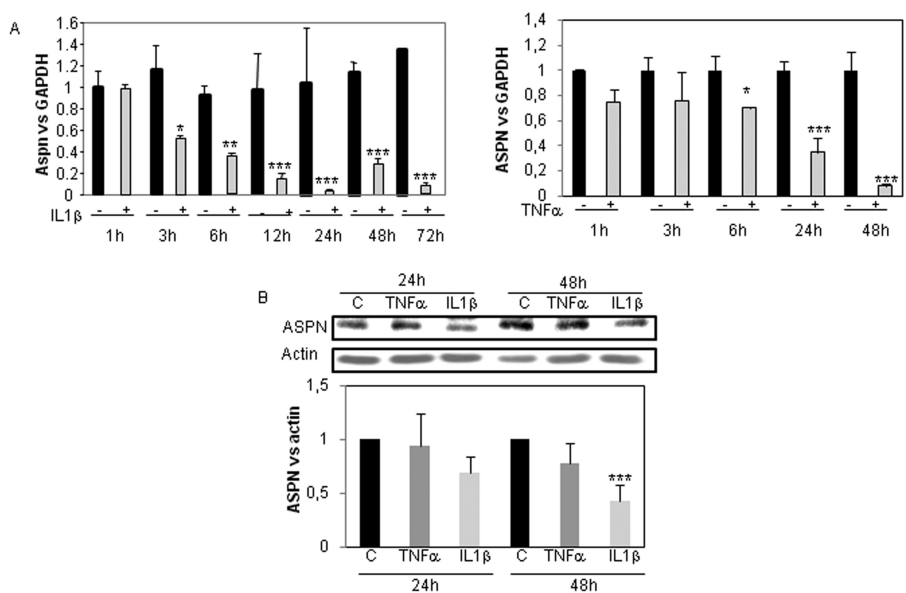

**Figure 2.** ASPN inhibition by IL-1β and TNFα (time-response). Primary cultures of HACs (P0 (no passage)) were cultured for 6–7 d in DMEM + 10% FCS and then maintained for 24 h in DMEM + 2% FCS before adding 1 ng/mL IL-1β or TNFα. They were incubated with these cytokines for the indicated times. ASPN mRNA level was defined by RT-PCR. Graphs represent modulation of ASPN mRNA level after normalization to GAPDH signal. Values are the mean and SD of triplicate experiments (A). Similar experiments were performed before protein extraction. Then ASPN protein level was analyzed by Western blot. The amounts of ASPN protein were quantified by densitometry using ImageJ, normalized against β-actin. Histograms represent the mean of values relative to control and SD of three experiments (B). C, control.

### Transfection Experiments

Sp1 expression vector (pEVR2-Sp1) was obtained from Guntram Suske (Institut fur Molekularbiologie and Tumorforschung, Marburg, Germany). ASPN-luciferase fusion plasmids were previously described (10). Chondrocytes were transiently nucleofected as previously described (11). After overnight transfection, cells were incubated for 24 h in DMEM containing 2% FCS.

### Luciferase Assay

Cells were washed once with PBS and harvested in lysis buffer. Luciferase activity was assayed on total cell extracts (Luciferase Assay Kit; Promega, Charbonnières, France) in a luminometer (Berthold Centro LB 960; Berthold, Thoiry, France), and β-galactosidase activity was assayed by a colorimetric assay. Luciferase activities were normalized to transfection efficiency, and transcriptional activities were expressed as relative luciferase units (mean ± SD of triplicates).

### Nuclear Extracts and Electrophoretic Mobility Shift Assay

Nuclear extracts were prepared as described previously (11). For electrophoretic mobility shift assay (EMSA), 5 fentomoles [γ-$^{32}$P]ATP-labeled probes and 7 μg of nuclear extracts from chondrocytes were incubated at room temperature for 15 min in binding buffer (20 mmol/L Tris-HCl, pH 7.5, 1 mmol/L dithiothreitol, 100 mmol/L NaCl, 5% bovine serum albumin, 0.1% Nonidet

P-40, 10% glycerol, 25 μmol/L $ZnCl_2$) and 2 μg poly(dI-dC)·(dI-dC). The following probes were used (only the forward strand is shown): AGTTTCCCTG ACATGCTATAGCAG (probe 1); AAAAAAAGTATGGTTCTT (probe 2); CATCTGAAGTTTGGGTATATAC (probe 3); GTCTATACAGGTTGAGTAAT TTTT (probe 4); and CTCCGACTGC ACTTTCAATGGCCAG (probe 5). The putative Sp1 binding site is underlined. Samples were resolved on a 7% nondenaturing polyacrylamide gel and exposed for autoradiography.

### Statistical Analysis

All experiments were repeated at least 3× each with different donors, and similar results were obtained. Only representative experiments are shown. Data are presented as the mean ± SD of triplicate experiments. Statistical significance was determined by the Student $t$ test. $P$ values <0.05 were considered significant (***$P$ < 0.001; **$P$ < 0.01; *$P$ < 0.05 versus controls).

## RESULTS

### Proinflammatory Cytokines Downregulate Asporin Expression

We treated human articular chondrocytes with IL-1β and TNFα. These treatments mimic, *in vitro*, inflammatory aspects observed during OA process; in particular, they strongly induce expression of metalloproteinases (MMP1, MMP3, MMP9 and MMP13) and cytokines (IL-6 and IL-8) (Figure 1).

First, we examined the effect of IL-1β and TNFα (1 ng/mL) on asporin expression in human articular chondrocytes cultured in DMEM + 2% FCS. Using real-time reverse transcription–polymerase chain reaction (RT-PCR) (Figure 2A), we showed that these proinflammatory cytokines significantly reduce asporin mRNA level. This marked decrease was seen as early as 3 and 6 h of incubation with IL-1β and TNFα, respectively, and was maintained for at least 48 h. At the protein level, 48 h of treatment with IL-1β reduced ASPN expression, whereas 24 h of incubation did not induce a significant effect. In addition, we were unable to detect any effect after both 24 and 48 h of treatment with 1 ng/mL TNFα (Figure 2B).

Dose-response experiments showed that 0.1 ng/mL IL-1β for 48 h is sufficient to repress asporin mRNA expression (70%) (Figure 3A), whereas a higher concentration (>1 ng/mL) was required to reduce protein expression (Figure 3B). Similarly, 0.5 ng/mL TNFα was sufficient to drop ASPN mRNA expression, whereas at least a 10-fold higher concentration was required to induce reduction of ASPN protein expression (Figure 3).

### TGFβ Differentially Modulates Asporin Expression according to FCS Amounts

We pursued this study by the analysis of TGFβ1 effect on ASPN (Figure 4). We treated cells for increased times with 5 ng/mL TGFβ1 or with increased doses for 24 h in DMEM + 2% FCS. Surprisingly, we obtained no significant modulation of ASPN mRNA level upon TGFβ1 treatment (Figure 4A, B). This is in disagreement with Igekawa's work, which observed an increase under TGFβ (10). But, these authors treated cells in 0.2% FCS. For this reason, we reproduced this experiment with a different concentration of FCS (0% to 2% to 10%) (Figure 4C). First, we found that asporin mRNA level decreased with the amount of FCS. Second, as in the preceding figure, TGFβ did not affect asporin mRNA level when chondrocytes were cultured in the presence of FCS (2% or 10%). However, in FCS-free medium, TGFβ significantly increased asporin expression. This induction was also observed at the protein level (Figure 4D).

### Dedifferentiation of HAC Reduces ASPN Expression

ASPN expression was evaluated at mRNA levels in chondrocytes that un-derwent up to three passages before inclusion in the three-dimensional scaffold (alginate). These successive passages lead dedifferentiation of chondrocytes characterized, among others, by a low level of collagen type II and aggrecan and a high expression of collagen type I and III (Figure 5). On the contrary, alginate culture of chondrocytes passaged three times restores, at least in part, the phenotype of chondrocytes with an increased expression of collagen type II and a reduction of collagen types I and III (see Figure 5) (8). Clearly, whereas ASPN expression decreased in a passage-dependent manner (Figure 6A), it strongly increased when chondrocytes, dedifferentiated after three passages, were cultured for 3 wks in alginate beads (Figure 6B).

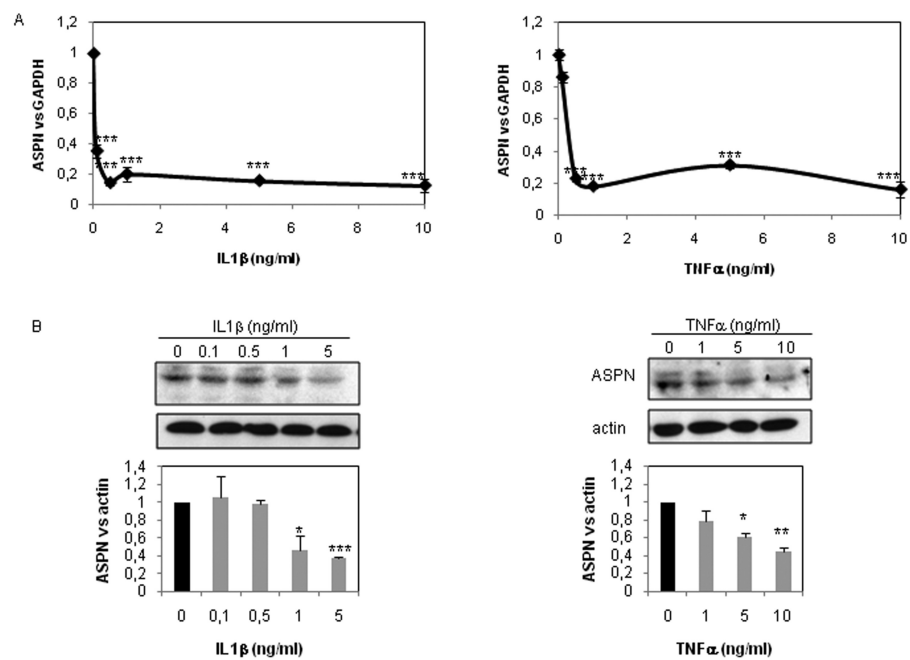

**Figure 3.** ASPN inhibition by IL-1β and TNFα (dose-response). HACs were incubated with increased dose of IL-1β and TNFα for 48 h. ASPN mRNA level was defined by RT-PCR. Graphs represent modulation of ASPN mRNA level after normalization to GAPDH signal. Values are the mean and SD of triplicate experiments (A). Similar experiments were performed before protein extraction. Then ASPN protein level was analyzed by Western blot. Histograms represent the mean of values relative to control and SD of three experiments (B). ***$P$ < 0.001; **$P$ < 0.01; *$P$ < 0.05 versus controls.

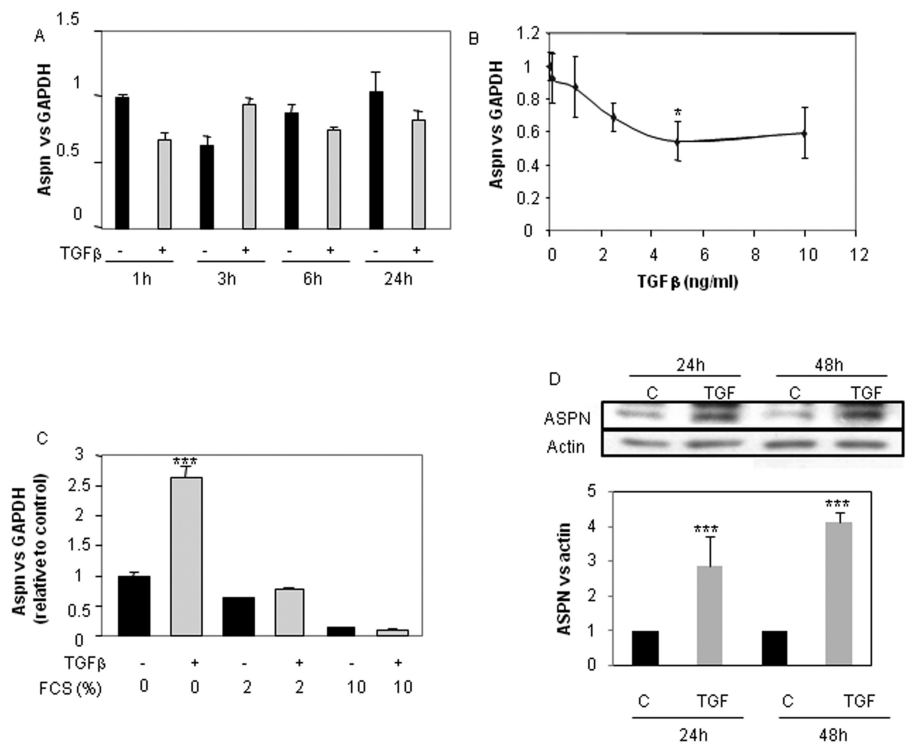

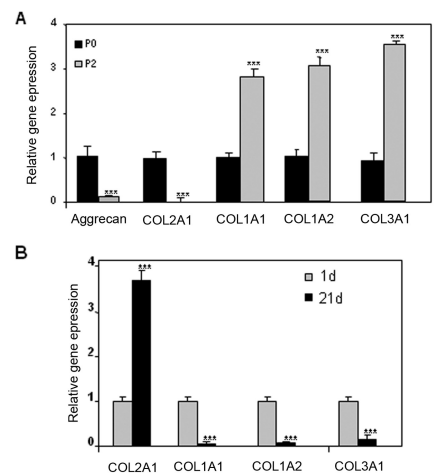

**Figure 4.** Differential regulation of ASPN expression by TGFβ. HACs were cultured as in Figure 1. Then, after a 24-h incubation in DMEM + 2% FCS, they were treated with 5 ng/mL TGFβ for the indicated times (A) or with increased concentration for 24 h (B). HACs were also incubated in 0, 2 or 10% FCS for 24 h before adding TGFβ (5 ng/mL) for an additional 24-h period (C). ASPN mRNA expression was defined by real-time PCR. HACs were also treated with TGFβ1 for 24 or 48 h in the absence of serum, and Western blot was performed. Histograms represent the mean of values relative to control and SD of three experiments (D). ***$P < 0.001$; **$P < 0.01$; *$P < 0.05$ versus controls. C, control.

**Figure 5.** Passages induce alteration of HAC phenotype, which is reversed by alginate culture. Human articular chondrocytes were passaged twice and subsequently maintained in culture for 7 d. At the end of the incubation, RNA was extracted and reversed transcribed. Then, aggrecan, COL2A1, COL1A1, COL1A2 and COL3A1 mRNA levels were assayed by real-time PCR (A). P3 (at three passages) HACs were also cultured 1 or 21 d in alginate beads (B–D,) and collagen mRNA levels were assayed by real-time PCR (B). ***$P < 0.001$; **$P < 0.01$; *$P < 0.05$ versus controls. P3, at three passages.

### Sp1 Upregulates ASPN Expression through the –473/–140 Region of the ASPN Promoter

Sp1 is one of the major transcription factors controlling chondrocyte metabolism. Indeed, it regulates numerous matrix genes, such as collagen II, decorin and biglycan (12,13). Because Sp1 is downregulated during dedifferentiation of chondrocytes (Figure 7) (14) and upregulated by alginate culture (see Figure 5), we hypothesized that this transcription factor may regulate ASPN expression. Figure 7A shows that Sp1 ectopic expression increased ASPN expression in both primary and twice-passaged chondrocytes.

Then, we wondered whether Sp1 was able to bind to the ASPN promoter. First, transient cotransfection of ASPN promoter and Sp1 expression vector revealed that Sp1 increases the transcriptional activity of a construct driven by the –473/+14 ASPN promoter in both primary and passaged chondrocytes. On the contrary, Sp1 was not able to increase activity of a shorter construct (–140/+14), suggesting that Sp1 may act through a region located between –473 and –140 bp upstream of the transcription start site (Figure 7B, C).

### Sp1 Is Able to Bind to the –473/–140 Region of the Human ASPN Promoter

Analysis of this ASPN promoter (region –473/–140) with Patch_Search (http://www.gene-regulation.com/cgi-bin/pub/programs/patch/bin/patch.cgi), TFBind (15) and TRED (http://rulai.cshl.edu/cgi-bin/TRED/tred.cgi?process=searchTFSiteForm) showed five putative binding sites for Sp1 (Figure 8A) in this region. We used these identified sequences as probes in EMSA experiments to determine whether Sp1 is able to bind to the ASPN promoter. These probes were incubated with nuclear extracts of primary HACs transfected with pEVR2-Sp1 or pEVR2. One delayed protein-DNA complex was observed with probe 4 (–238/–261) and with a weaker intensity with probes 1 (–452/–428) and 5 (–142/–166). This complex was enhanced by ectopic expression of Sp1 for these three probes (Figure 8B). No bands were detected with probes 2 (–356/–339) and 3 (–330/–308) (not shown). Thereafter, antibody interference assays were performed with antibodies targeting Sp1, using probe 4. The binding activity of the complex to probe 4 was decreased in the presence of the Sp1 antibody, indicating that this complex contains Sp1 (Figure 8C). Therefore, Sp1 is able to bind to the –473/–140 region of the human ASPN promoter.

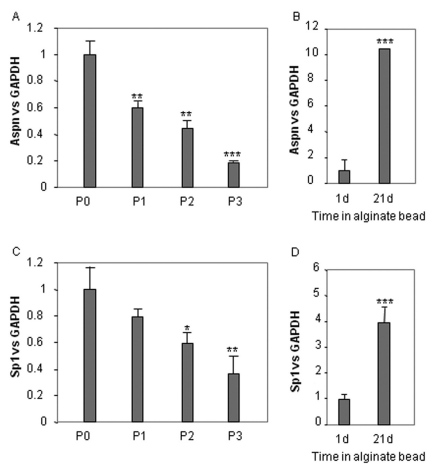

**Figure 6.** ASPN expression according to differentiation status of chondrocytes. Human articular chondrocytes, cultured as described in Materials and Methods, were passaged up to three times and subsequently maintained in culture for 7 d (A–C). P3 HACs were also cultured 1 or 21 d in alginate beads (B–D). At the end of the incubation, RNA was extracted and reversed-transcribed. Then, asporin and Sp1 mRNA level was assayed by real-time PCR. ***$P < 0.001$; **$P < 0.01$; *$P < 0.05$ versus controls. P3, at three passages.

## DISCUSSION

Asporin has been reported to be abundantly expressed in OA articular cartilage (2). However, few data have been reported about its regulation. In this study, we described the regulation of asporin expression in human articular chondrocytes. Our data demonstrate that asporin expression is regulated by the major cytokines involved in cartilage homeostasis and OA, namely TGFβ, IL-1β and TNFα.

First, we tried to mimic the OA process *in vitro* in two different ways. On one hand, we treated HACs with IL-1β and TNFα to reproduce the increased levels of proinflammatory cytokines observed in OA synovial fluid. On the other hand, we analyzed ASPN expression in passaged HACs. Indeed, repeated passages of chondrocytes mimics some of the alterations of the chondrocyte phenotype observed during OA, at least in the late stages of the disease (stage of fibrocartilages). Passages induce morphogenic alterations of cells, accompanied by profound biochemical changes including the loss of production of cartilage-specific macromolecules, that is, type II collagen in the profit of type I collagen synthesis (16). However, this model is not able to reproduce the other modifications undergone by OA chondrocytes and notably the induction of type × collagen, specific of hypertrophy stage. Thereafter, we used the alginate bead, a culture system that has been shown to revert to a chondrocytic profile from a phenotypically altered chondrocytes by several passages on plastic flasks (17,18).

We found that proinflammatory cytokines, as well as chondrocyte dedifferentiation, decrease ASPN expression. This ASPN downregulation was surprising. Indeed, analysis of microarrays publicly available in Gene Expression Omnibus (NCBI databases) suggests that ASPN expression is induced during OA (GSE8077) and rheumatoid arthritis (GDS2952). However, a similar regulation was reported for other close small leucine-rich proteoglycans. Whereas decorin and biglycan concentrations are higher in OA cartilage (either in human OA [20,21] or in experimental models [20]), they are downregulated by IL-1β and passages (22,23).

On the contrary, ASPN expression is upregulated by TGFβ1. Similar results with BMP-2/TGFβ were already recently reported (10,24). However, we answer questions raised by these reports, since we demonstrate here for the first time that TGFβ exerts its positive effect on ASPN only when cells are cultured in the absence (our data) or with very weak levels of FCS (0.2% [10]). Further experiments would be required to understand the mechanism involved in the differential effect of TGFβ according to FCS level. However, a different hypothesis may be proposed. First, several factors contained in FCS modulate biological effects of TGFβ. For instance, fibroblast growth factor (FGF) acts synergically with TGFβ to inhibit collagen II synthesis in chondrocytes (25). On the contrary, other proteins contained in the serum, such as α2-macroglobuline, are able to sequestrate TGFβ and block its interaction with its receptors and subsequently its effects. Another hypothesis is ASPN expression may depend on the proliferation state of chondrocytes. Indeed, TGFβ1 differentially regulates type II collagen expression in proliferative and nonproliferative chondrocytes (26).

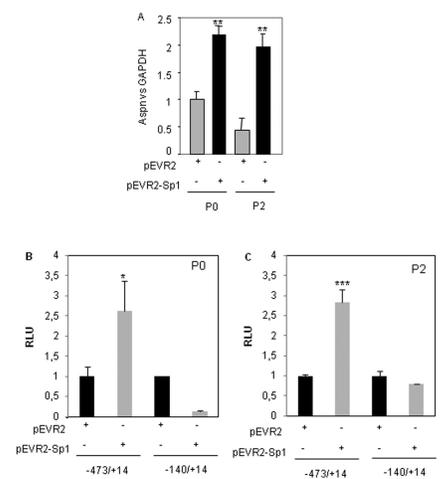

**Figure 7.** Sp1 upregulates ASPN expression in chondrocytes. HACs (P0 (no passage) or P2 (at two passages)) were transfected with Sp1 expression vector (pEVR2-Sp1) or with the insertless plasmid (2 μg). After 6 h, the medium was changed to DMEM + 2% FCS for 24 h. Total RNA was isolated and asporin and Sp1 mRNA levels were analyzed and normalized with GAPDH level. Values are the mean and SD of triplicate experiments (A). P0 or P2 HACs were nucleofected with 2 mg pEVR2-Sp1 (or insertless pEVR2 vector) and ASPN promoter constructs. After 24 h, luciferase activities (relative luciferase units (RLU)) were determined and expressed as percent of controls (B, C). ***$P < 0.001$; **$P < 0.01$; *$P < 0.05$ versus controls.

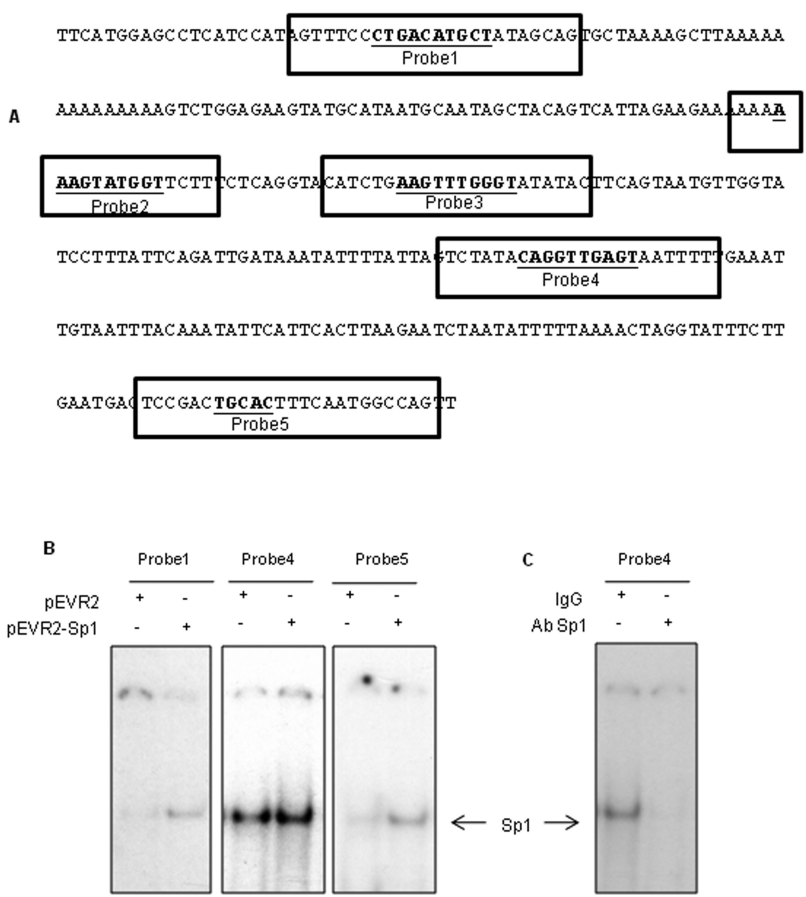

**Figure 8.** Sp1 binds to the ASPN promoter. The region –473/–140 of the human *ASPN* gene was analyzed with Patch_Search, TFBind and TRED. The putative Sp1 DNA binding sites are in bold (A). EMSA reactions were performed using the 5′ end–labeled probes, incubated with 7 mg of nuclear extract from chondrocytes nucleofected with Sp1 expression vector (B). Nuclear extracts were also preincubated with IgG or Sp1 antibody before incubation with probe 4 (C).

The findings of Kou *et al.* (10) indicate that TGFβ1 induces *ASPN* through the Smad pathway and involves Smad3 in particular. They suggest that ASPN is the indirect target of Smad3, and hence, its expression could be influenced by other mediators, such a *GADD45*, a factor that upregulates biglycan upon TGFβ treatment and is a target gene of Smad3 (10). In this study, we propose a role for Sp1, a transcriptional factor able to regulate decorin and biglycan expression. Here, we found that Sp1 upregulates ASPN expression and is able to bind to the ASPN promoter. Others transcriptional factors, especially Sox9, Ets and Runx family factors, may be hypothesized to be important in the control of ASPN expression. Indeed, interestingly, numerous putative binding sites for Hox/Runx and Ets/FKHD/Stat are present in the first intron of ASPN (27). Therefore, because Smads are known to modulate transcription of Runx and HOX transcription factors, whereas IL-1β induces Ets family expression, these factors may be potential mediators of *ASPN* expression.

In summary, the present work documents the regulation of ASPN expression in human articular chondrocytes and enlightens the crucial influence of cytokines and cell differentiation status. However, while we present new knowledge in OA physiopathology, it is difficult to conciliate our *in vitro* findings with data obtained from OA patients. Nevertheless, it seems that this apparent contradiction between *in vitro* results and *in vivo* observations is general to class I SLRP and that ASPN is regulated in the same way as decorin and biglycan.

## ACKNOWLEDGMENTS

We thank Dr. Suske (Institut fur Molekularbiologie and Tumorforschung, Marburg, Germany) for providing pEVR2-Sp1 and Thomas Osouf for his help with the analysis microarray. E Duval, N Bigot and M Hervieu received fellowships from the Regional Council of Lower Normandy. This work was supported in part by the Lions Clubs from Normandy (France).

## DISCLOSURE

The authors declare that they have no competing interests as defined by *Molecular Medicine*, or other interests that might be perceived to influence the results and discussion reported in this paper.